# Energy-efficient Non Uniform Last Level Caches for Chip-multiprocessors Based on Compression

Pooneh Safayenikoo, Arghavan Asad, and Mahmood Fathy

**Abstract**— With technology scaling, the size of cache systems in chip-multiprocessors (CMPs) has been dramatically increased to efficiently store and manipulate a large amount of data in future applications and decrease the gap between cores and off-chip memory accesses. For future CMPs architecting, 3D stacking of LLCs has been recently introduced as a new methodology to combat to performance challenges of 2D integration and the memory wall. However, the 3D design of SRAM LLCs has made the thermal problem even more severe. It, therefore, incurs more leakage energy consumption than conventional SRAM cache architectures in 2Ds due to dense integration. In this paper, we propose two different architectures that exploit the data compression to reduce the energy of LLC and interconnects in 3D-ICs. The fundamental idea of the proposed architectures is to use the non-uniform distribution of the accesses, invalid lines, zero–value lines, and frequent-value (FV) lines in banks of an LLC for decreasing both dynamic and leakage energy. The proposed architectures based on non-uniform cache architectures (NUCA) disable cache banks with low accesses and high invalid, zero, and FV lines, leading to high energy efficiency.

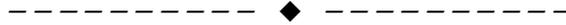

## 1 INTRODUCTION

With the development of semiconductor technology, on-chip transistor densities have increased steadily, enabling the integration of many cores on a single die. However, all fabricated transistors on a silicon die cannot be simultaneously powered on or utilized. This phenomenon has been termed dark silicon [1]. Today, how to leverage these dark or idle transistors is the main challenge to improve energy and performance. This challenge gave rise to creative methods on power-efficient architectures. Therefore, it is essential to provide next-generation architectural techniques, design tools, and analytical models for future many-core Chip-Multiprocessors (CMPs) in the presence of dark silicon [2].

Most of the prior works in the dark silicon [1] [3] [4] are characterization studies and focus on cores rather than uncore components. However, the challenges of dark silicon in-memory architecture are critical for computational applications and applications with the high locality that the memory hierarchy is dark most of the time. In this work, we decide to answer this question: Is there any architecture that explores the role of compression in the dark silicon era? Simply put, is there an advantage in storing compressed data in the traditional SRAM cache banks to power off? In this work, we are going to respond to this question by monitoring the type of data, the distribution of them in cache banks, and access to them

On the other hand, an emerging technology in many-core systems in three-dimensional integrated circuits built by vertical stacking of dies and uses Through-Silicon-Vias (TSVs) to connect the different layers. This method reduces interconnection wire length, lowering power consumption and shortening communication latency [5]. Nevertheless, the capacitive crosstalk, which is one of the significant contributors to delay and power consumption of on-chip interconnect, is not omitted in 3D designs and may degrade the performance of TSVs. Crosstalk noise can also affect chip timing by slowing down transitions on a switching signal when a large number of its neighbors perform opposite transitions. Therefore, interconnects in 3D are a significant contribution to system power consumption. In addition, they are vital limiters of system performance [6] [7].

Due to short-channel effects (SCEs) in nano-scale technology, leakage power depletes the power budget and substantially contributes to overall power consumption. Although FinFET as an emerging technology has significantly reduced leakage power due to better gate control [8], power efficiency is a relative concept. Therefore, the leakage reduction method is still used in the FinFET technology context.

Cache systems are among the most power-hungry components in multi/many-core CMPs. The leakage power within them has become a significant contributor to nanoscale technology's overall chip power budget [9]. Thus, architecting an energy-efficient memory hierarchy with minimum leakage energy is critical for embedded systems.

In recent years, to address the requirement of new data-centric applications for more prominent memories on the one hand, and to break the memory wall problem, on the other hand, architects have dramatically increased the size of the last-level cache (LLC) in the memory hierarchy.

---

• P. Safayenikoo is with the Computer Science Department of University of California Santa Cruz, Santa Cruz, CA, USA. E-mail: psafayen@ucsc.edu.

Non-uniform cache architectures (NUCAs) occur because the caches are so large that the latencies to different parts are different [10]. Therefore, these architectures have been introduced as the primary cache organizations for future many-cores. To improve the energy efficiency of CMPs with NUCA caches, this paper proposes two architectures that attempt to power off cache slices placed in tiles. These techniques are used when the cores and LLC of a microprocessor are implemented on different dies, and a dense 2D mesh of TSVs is used to connect them. This coding is applied at run-time to each tile.

In the first design, known as NIZCache, we focus on a common form of compressible data, cache lines that contain only zero as values, and aggregates (entire banks) that are predominately full of zero values or invalid lines. Therefore, a bank in NIZCache is determined power-off or power-on at runtime by estimating these three parameters (access, invalid, and zero lines) for reducing leakage energy. This technique uses power gating at the granularity of the bank.

In the second design, called NFVCache, we propose another new architecture based on a constant length compression in uncore components to reduce the crosstalk delay and leakage energy. To these ends, we first find a few data moved in uncore components repeatedly by using a static profiling method. Then we encode these data by Limited-Weight Code (1-LWC) [11]. We adopt frequent values based on encoding to optimize power in uncore components. Hence, we can disable the banks that have high frequent value counts related to access counts for a specific time.

In this paper, we make the following contributions:
- We propose two new architectures based on a constant length compression in uncore components to reduce the uncore energy, cache hierarchy, and on-chip interconnect that significantly contribute to the overall chip power budget in the nanoscale era. Our proposed architectures use a well-known dynamic power management technique (i.e., compression) to reduce the crosstalk delay and leakage energy simultaneously, as the first time.
- A new statistical approach is proposed to find the non-uniform distribution of accesses across the cache banks. We use statistical information instead of specific thresholds to distinguish banks in the non-uniform cache architecture. Therefore, since the statistic-based power-off policy is decided individually based on the program behavior, it increases the miss rate less than the threshold-based power-off policy. Therefore, NIZCache and NFVCache can benefit from the strength of this statistical information with less performance degradation. Furthermore, a new power-on approach is proposed that is considered turning on each bank individually along with turning on several banks. Therefore, this method determines which bank is the best choice for shutting on.
- We exploit the type of data such as zero data and frequent values (few data that are moved in uncore components repeatedly) to power off banks as the first work. Due to frequent-value locality and high spatial locality of null cache lines, we can use the opportunity of these data to decrease the overhead of migration and improve the miss rate.
- We evaluate energy-saving and performance of our designs compared to EECache [12]. The results of NIZCache represent 17% and 14% energy-delay product improvement on average compared to EECache under multithreaded and multi-programmed workloads, respectively. On the other hand, the results for NFVCache show that this design improves the energy-delay product by about 26% and 25%, on average compared to EECache under multithreaded and multi-programmed workloads, respectively. We also compare the energy consumption of our work with the state-of-the-art compression-based cache design [27], and the result of NFVCache presents a 22% energy improvement.

The rest of this paper is organized as follows: Section 2 gives essential background and the motivation for our work. A brief background on crosstalk in TSVs and LWC coding is explained in Section 3. Section 4 explains the proposed runtime methods. Section 5 presents the experimental setup followed by experimental results. Section 6 describes the related work. Finally, the paper concluded in Section.

## 2 BACKGROUND AND MOTIVATION

### 2.1 Data Organization in Last Level Cache

In this work, we propose an energy-efficient LLC architecture. Therefore, we focus on the top layer of a 3D CMP with a stacked cache. TSVs connect two processing layers in this architecture. The bottom layer consists of cores and private L1 caches, and the second-level cache has 64 banks and is placed on the top layer. Our proposed LLC architecture is based on dynamic non-uniform cache architectures. In this NUCA, LLC cache banks are connected by a mesh interconnection network, and the address space is partitioned across cache banks. We present a simulation result to compare the distribution of accesses. Note that Section 5 shows the system configuration in detail.

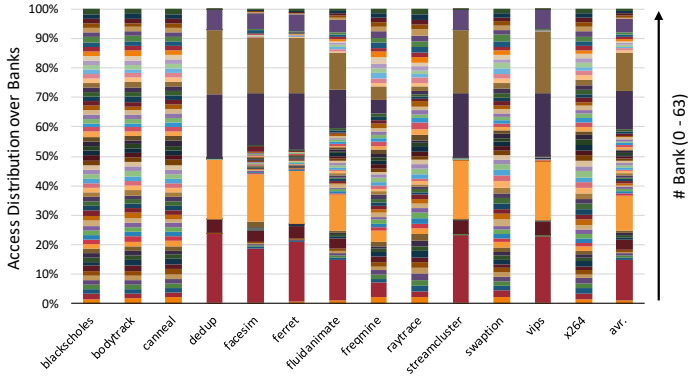

Fig. 1. The percentage of accesses related to the L2 cache banks.

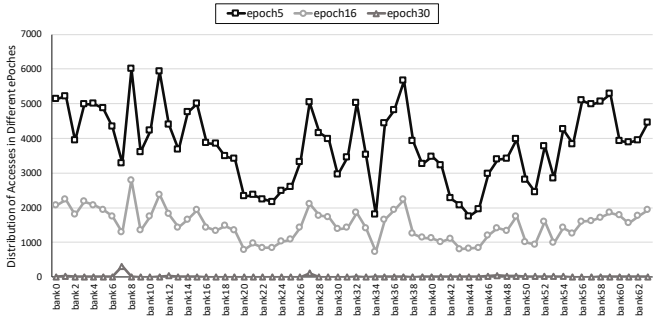

Fig. 2. The number of accesses of L2 cache banks in raytrace benchmark.

Due to the spatial locality of some programs, a bank can service a chain of requests that are caused to uneven capacity pressures. Generally, this point is shown in Fig. 1 for all PARSEC benchmarks [13] in the mentioned architecture. Due to this histogram, a few cache banks cover most accesses to the L2 cache. For example, in *vips* and *streamcluster* benchmarks, only six banks occupy 95% and 97% of total accesses, respectively. Furthermore, the number of accesses may also vary in different program periods. Fig. 2 shows the number of accesses over the L2 cache banks in the *x264* benchmark for three intervals (each epoch is 64Mcycles). As can see, the access counts across L2 cache banks are non-uniform in various program phases. These observations motivate us to use the uneven accesses to design the last-level cache. In addition, we decrease this contention significantly by using encoding.

### 2.2 Quantifying Data in Last Level Cache
#### 2.2.1 Zero Data
On the other hand, it has been shown that large amounts of data in many applications are zero [14]. Therefore, the cache space is wasted by storing zero data blocks in the memory hierarchy. Thus, it can be helpful if zero data access indicates a significant part of cache accesses. Table 1 shows the access to zero-valued lines in the L2 cache for one billion instructions on PARSEC benchmarks. We use this abundance to reduce the energy of accesses by writing and reading a single bit for every zero line and reduce the miss rate in power-off banks.

Furthermore, these zero data frequently have high spatial locality. We show the percentage of zero lines in data cache accesses over banks in Fig. 3. This figure shows that 64% of zero lines access only two banks for the *swaption* benchmark. This locality of zero lines provides an opportunity to power off since when these banks turn off, it will cause a few additional cache misses and an insignificant loss of performance.

TABLE 1
FREQUENT VALUE ACCESS PER KILO-INSTRUCTION (FVAPKI), NULL BLOCK ACCESS PER KILO-INSTRUCTION (NAPKI), AND ACCESS PER KILO-INSTRUCTION (APKI) ON A 32MB L2 CACHE

| Application | FVAPKI | NAPKI | APKI |
|---|---|---|---|
| Blackscholes | 22.975665 | 1.75451 | 40.35713 |
| Bodytrack | 0.934211 | 0.43862 | 128.662 |
| Canneal | 1.336622 | 0.033281 | 80.99452 |
| Dedup | 33.502939 | 0.051509 | 104.4628 |
| Facesim | 24.871853 | 4.244459 | 79.34748 |
| Ferret | 24.053897 | 4.318626 | 82.00954 |
| Fluidanimate | 5.18144 | 2.505694 | 27.04751 |
| Freqmine | 0.342655 | 0.256463 | 1.508276 |
| Raytrace | 0.499373 | 0.291605 | 5.599993 |
| streamcluster | 32.944733 | 0.164743 | 103.2058 |
| Swaption | 0.067315 | 0.065192 | 0.294473 |
| Vips | 34.272193 | 0.348685 | 109.5816 |
| x264 | 1.190145 | 0.803034 | 25.44552 |

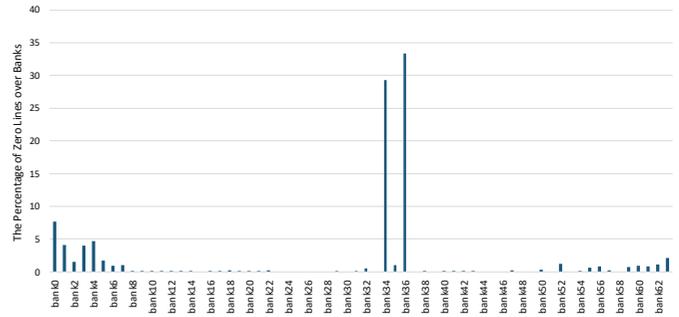

Fig. 3. The distribution of zero lines crosses the L2 cache banks in the swaption benchmark.

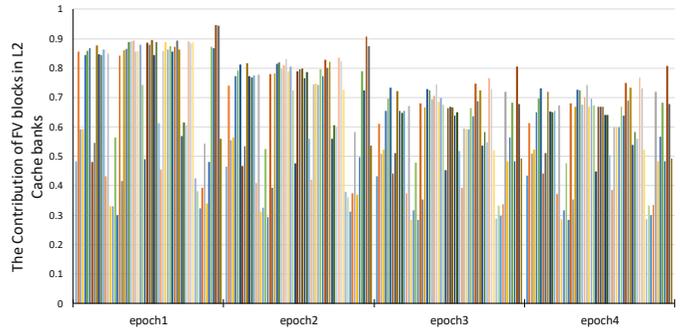

Fig. 4. The distribution of frequent value lines across the L2 cache banks in four epochs for MB1.

Due to these points, it is efficient to disable banks with higher zero lines, and indeed, it provides an opportunity to power off banks. Therefore, our design uses zero data to reduce the dynamic energy and non-uniformity zero lines over cache banks to reduce the leakage energy.

#### 2.2.2 Frequent Value Data
Several distinct values appear more frequently at any given point in the program execution than the observed ones. These values are called frequent values (FV), and this phenomenon is known as frequent values locality

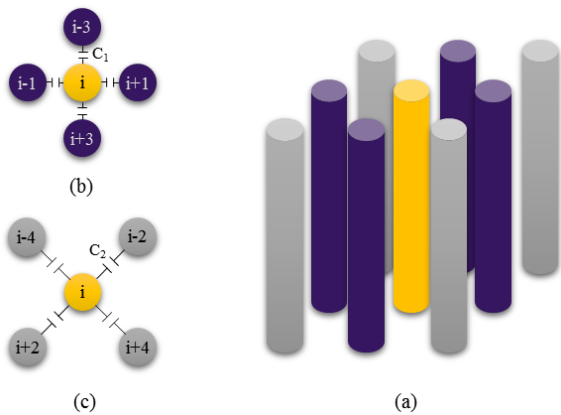

Fig. 5. A 3D display of a TSV bundle beside the capacitance crosstalk model.

[15]. Some researchers used this locality to store data in compressed form. We conducted an experiment based on the mentioned configuration to observe the frequent value locality in real applications. In this experiment, we first investigated the Frequent Value Access per Kilo-Instructions (FVAPKI) in the PARSEC suite as our benchmarks while the number of FV is 32, and the bit length FV is 64 bytes. Therefore, we can see a high frequent value locality in the LLC. Then, we calculated the distribution of FVs across L2 cache banks. The results showed that this distribution is not uniform. A large percentage of accesses to banks; for example, the distribution of frequent value lines across the L2 cache banks in four consecutive intervals for the *MB1* benchmark is shown in Fig.4. We can also observe that this frequency does not usually change over time. Consequently, this observation motivates us to disable a significant part of these banks by using compression. Therefore, disabling these banks provides a different opportunity for powering off and significantly decreases the miss rate.

## 3 PRELIMINARIES

### 3.1 Crosstalk in TSVs

Fig. 5 shows the square model of the TSV in a 3D view where other TSVs surround a TSV. The TSVs can be categorized into two distinct sets based on their center-to-center distance: the TSVs placed vertically/horizontally and the diagonal TSVs. We use the RC model to examine the crosstalk effect on circuit delay (Fig. 5 b and c). Due to the definition of capacitance and the distance between vertical/horizontal TSVs and diagonal TSVs from the center TSV, the coupling capacitance between wires is different. As can be seen, we use $C_1$ to present the coupling capacitance of vertical/horizontal TSVs, and $C_2$ to give the coupling capacitance of diagonal TSVs. Hence, the approximation delay under the crosstalk effect can be calculated as follows [20]:

$$\tau_i(\alpha) = -R\left(C + C_1(\delta_{i,i-3} + \delta_{i,i-1} + \delta_{i,i+1} + \delta_{i,i+3}) + C_2(\delta_{i,i-4} + \delta_{i,i-2} + \delta_{i,i+2} + \delta_{i,i+4})\right)$$

$$\Delta V = V(t^+) - V(t^-)$$
$$\delta_{i,j} = abs\left(\frac{\Delta V_i - \Delta V_j}{V_{dd}}\right) \quad (1)$$

In this equation, C is the capacitance between TSV and the substrate, and $V_i(t+)$ and $V_i(t-)$ represent the voltages before and after the transition. Due to the signal transition direction between the center TSV and its neighbors, the value of $\delta_{i,j}$ can only be 0, 1, or 2. For example, if the signal in the center TSV is switching from logic 0 to logic 1 and also signal in a neighboring TSV is switching in the opposite direction, $\delta_{i,j}$ takes the value of 2. If the two signals switch in the same direction over, $\delta_{i,j}$ equals 0. Note that there is no signal delay when signal i does not change. Finally, the effective our crosstalk capacitance on TSV, if the ratio between $C_1$ and $C_2$ equals to 0.5, i defined as follows:

$$C_{eff,i} = \left(C + C_1(\delta_{i,i-3} + \delta_{i,i-1} + \delta_{i,i+1} + \delta_{i,i+3}) + C_1/4(\delta_{i,i-4} + \delta_{i,i-2} + \delta_{i,i+2} + \delta_{i,i+4})\right) \quad (2)$$

### 3.2 Light Weight Coding (LWC) Encoding

In order to minimize both energy and crosstalk, we present the Light Weight Coding (LWC) encoding mechanism. The weight of a data is defined as the number of 1s in that data. An m-Limited Weighted Code is a coding mechanism in which the weight of codewords is less than or equal to *m* [21]. Indeed, the maximum weight is limited in the codeword. Therefore, the number of transitions is limited by the maximum weight. We use 1-LWC codes in this paper for 32 frequent values. We encoded 64-byte frequent values with 32-bit that has only one bit 1, and this bit for each FV is in different places in the codeword. Therefore, the crosstalk is eliminated since no adjacent 1s in the codewords. On the other hand, we can use 60 bytes free caused by coding in blocks of L2 cache banks to power-off opportunities and thus the leakage energy reduction.

## 4 OUR PROPOSED METHODS

This section two proposed cache organizations and their power off and power on policies, respectively. This section describes the cache controller to determine the power-on or the power-off banks in the LLC at each interval. Generally:

- First, we monitor the cache behavior consisting of the number of zero-value lines (just in NIZCache architecture), frequent values (just in NFVCache architecture), the number of accesses, and invalid lines in each bank at the beginning of each interval execution.
- Second, at the end of each interval, using the gathered monitoring information, we will decide if there is an opportunity to turn off banks of LLC for the next break.

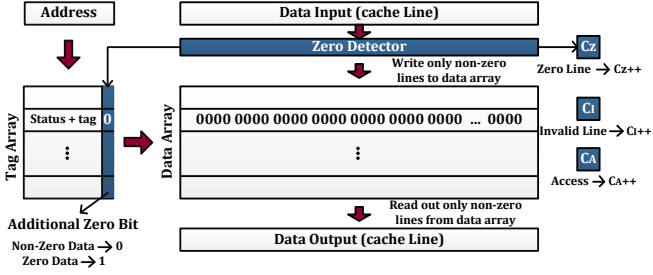

Fig. 6. Organization of an L2 cache bank.

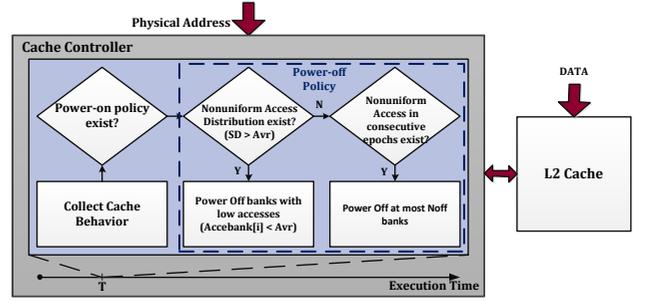

Fig. 7. Overview of the statistic-based power-off policy.

- Third, if there are power-off opportunities for some banks, these banks are turned off in the next interval. Nevertheless, all counters keep turning on to collect the information of banks to change the state of the banks due to the changes of workloads in various periods. Therefore, power-on policies simultaneously are checked for power-off banks.

### 4.1 NIZCache

In this subsection, we propose a runtime cache architecture called NIZCache. The fundamental idea of NIZCache is to use the non-uniform distribution of the accesses, invalid lines, and zero–value lines. Extras in banks of the last level caches for decreasing dynamic and leakage energy. Fig. 6 shows an organization of one bank of NIZCache design. We use an array to place an extra bit at a granularity of blocks or lines to reduce dynamic energy and find an opportunity to disable SRAM banks by exploiting the prevalence of zero blocks. In write operations, NIZCache first detects the 64-byte zero-valued data to be written to the cache without any attention to power-off or power-on of cache banks. If the zero-valued data are detected, corresponding additional zero-bit sets to 0 in its tag array. Thus, only non-zero-valued lines are written to the power-on SRAM banks. In read operations, the NIZCache can readouts zero lines in both the power-on and power-off cache banks and only non-zero lines of the power-on banks. Thus, our cache reduces the number of write and reparations in an LLC and reduces dynamic energy.

### 4.2 NFVCache

Fig.8 shows the other architecture that we propose in this paper. NFVCache is based on non-uniformity of accesses, frequent values, and invalid lines over LLC banks. Indeed, this method uses constant length compression 1-LWC to simultaneously target the high leakage power of SRAM LLC and the crosstalk problem of Through-Silicon-Vias (TSVs). This architecture identifies frequent values by using the static profiling method to encode these values using limited weight codes and, therefore, minimize energy and crosstalk in NoC. Thus, the valid data must be identified since either the encoded or the original data can be stored in a data block. Consequently, we use a bit for each 64-byte word named an FV bit. As seen in Fig. 8 (a), we use an array that every entry corresponds to a cache line. There are 16 cache controllers in the second layer, as shown in Fig. 8.

| Algorithm 1: Statistic-based Power-off Policy |
|---|
| 1. For each period { |
| 2.  $C_A=\{C_{A0},C_{A1},…,C_{A63}\}$ //The number of accesses for each bank. |
| 3.  $C_C=\{C_{C0},C_{C1},…,C_{C63}\}$ //The number of compressed lines for each bank. |
| 4.  $C_I=\{C_{I0},C_{I1},…,C_{I63}\}$ //The number of invalid lines for each bank. |
| 5. N=0; //The number of powered-off banks |
| 6. |
| 7.  For (i=0; i<64; i++) |
| 8.   $C_{Xi}=C_{Ai}-C_{Zi}-C_{Ii}$ |
| 9. Calculate Mean (µ) and Standard Deviation (σ) of $C_X$ |
| 10. |
| 11. If ($\sigma_X > \mu_X$) |
| 12.  //There is non-uniform accesses to banks |
| 13.  For (i=0; i<64; i++) |
| 14.   If ($C_{Xi} < \mu_X$) { |
| 15.    $T_i=0$; //The i-th bank is powered off |
| 16.    N++; } |
| 17. Else |
| 18.  //There is uniform accesses to banks |
| 19.  If ($\mu_X(j-2) > 2*\mu_X(j)$ && N < $N_{off}$){ |
| 20.  //There is non-uniform accesses in consecutive periods |
| 21.   Shut down maximum $N_{off}$ banks with ith lowest number of $C_{Xi}$, $T_i=0$.} |
| 22. |
| 23.  $\mu_X(j-2) = \mu_X(j-1)$; |
| 24.  $\mu_X(j-1) =\mu_X$; |
| 25. } |

Each cache controller monitors the behavior of four cache banks connected to it. Therefore, due to using three 12-bit counters ($C_A$, $C_C$, $C_I$) and two 1-bit fields (the T field specifies the power state of each bank, the M field specifies the type of power-off for migration) for each cache bank, there are 12 counters and eight 1-bit fields in each cache controller.

When we find the frequent values and the corresponding codewords, we send them to each tile in the first layer. Therefore, we need to use a content-addressable memory (CAM) set in the network interface of each tile to store a frequent value table consisting of the 32 FVs and corresponding codewords. As shown in Fig. 8 (b), the input of the multiplexer (MUX) is composed of the original data and the output of the FV table. The actual data and output of the FV table are fed into a MUX, which is controlled by an FV bit. First, the original data is searched in the FV table; if the data is available, the valid bit is set to 1, and the encoded data is selected from the multiplexer. However, due to the existing 128 TSVs between each tile and the top layer, NI uses 32 out of 128 wires for data transferring, and the rest are OFF. Otherwise, the FV bit equals 0, and the original data is transferred on all 128 wires.

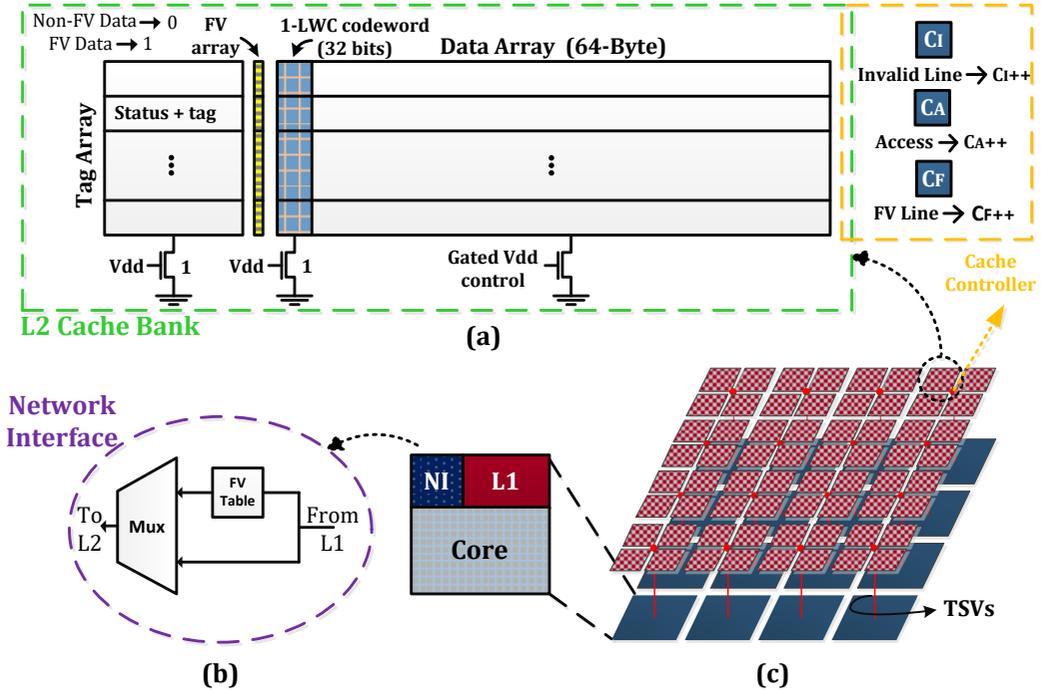

Fig. 8. (a) Organization of an L2 cache bank (b) Diagrams of the encoder placed in the network interface (c) Overview of the proposed NFVCache architecture.

The decoding process is very similar to the encoding process. First, the FV bit determines the number of ON wires to transfer data from the LLC cache. Therefore, if the FV bit equals 1, the FV table is active, and the decoded frequent value is chosen from the multiplexer. Otherwise, the FV table is inactive, and the original data is output from the multiplexer.

## 4.3 Power-off Policy

We propose two algorithms to disable unused cache banks. In each interval, all the fields and counters are reinitialized to 0. Then, the proposed approach decides how to power off unused SRAM banks in LLC by the gathered information from the monitoring procedure.

### 4.3.1 Statistic-based Policy

At each 64M cycle interval, the number of accesses, compressed lines, and invalid lines for each bank is counted. Then, at the end of each interval, the complete accesses to valid lines for each L2 cache bank are calculated at each cache controller ($C_{Xi}=C_{Ai}-C_{Zi}-C_{Ii}$). Each cache controller sends a control flit containing them to the monitoring unit. We consider one of the tiles in a core layer as a single monitoring unit that collects the complete accesses from the whole banks of the network. The communication between the monitoring unit and other cache controllers in the cache layer is handled by virtual point-to-point (VIP) connections [44]. Furthermore, the monitoring unit should be placed near the center to reduce the interconnection overhead between the cache controllers and the monitoring unit. Therefore this cache controller is in charge of conducting the reconfiguration process based on the mean and standard of the complete accesses of cache banks to determine the degree of the uniformity of accesses to different cache banks. When the distribution of accesses among the cache banks is uniform, the standard deviation value is small, vice versa [36]. For example, the *vips*, *streamcluster*, and *dedup* in Fig. 2 have a significant *standard deviation* for their $C_A$ array. After that, the following two conditions are checked in each period:

(i) If the standard deviation of complete accesses ($C_X$) in that interval is more than its mean, there is a non-uniform distribution of these accesses across banks. Banks whose $C_X$ are more than the mean complete accesses in that interval have the most accesses. Therefore, these banks are known as hot banks, and the rest are known as cold banks. These cold banks can be disabled to save leakage energy with negligible performance degradation in this state.

(ii) If there is uniform access to cache banks in that interval, we check the uniformity of accesses to cache banks in consecutive intervals. We compare the average complete accesses in three successive intervals to find nonuniformity. Therefore, if the mean complete accesses in two previous intervals are more than twice the mean complete accesses in the new interval and the number of powered-off banks is more minor than $N_{off}$, maximum $N_{off}$ banks are powered off. This number ($N_{off}$) depends on the architecture, and we set it to 16 empirically.

Otherwise, the distribution of complete accesses in LLC banks is approximately uniform in that interval compared to its two previous intervals. Thus, we do not change the state of SRAM banks. We present the overview of the proposed algorithm in algorithm 1 and Fig. 7.

EECache similarly uses three parameters, including utilization, hotness, and write-back of dirty data, to shut down the slices of the LLC for energy saving. In contrast, our work uses compressed, invalid, and several accesses to the cache banks. In both works, the number of powered-off banks should be less than a certain number ($N_{off}$) at each period, and some counters in the cache controller are added to determine the power state of cache banks. However, EECache decides based on the cache behavior while monitoring the cache behavior in consecutive periods. On the other hand, EECache is a threshold-based method that uses three specific thresholds to disable cache parts and requires high hardware overhead for sampling. At the same time, we propose a statistical-based approach that uses statistical information to distinguish banks in non-uniform cache architecture.

### 4.3.2 Threshold-based Policy

We disable a bank with low-access blocks that can be caused by high compressed lines in addition to low access counts to that bank. Therefore, this power-off policy selects the banks with a block-access count less than a certain threshold, $C_{th}$. The block-access count for active last level cache banks is calculated y:

$$BlockAccessCount_i = \frac{(A_i - C_i - I_i)}{\sum_{j=0}^{63} A_j} \quad (3)$$

The $A_i$, $C_i$, and $I_i$ in Equation 3 illustrate the number of accesses, compressed lines, and invalid lines of bank ith in the LLC, respectively. Therefore, when we have a threshold higher than $C_{th}$, the number of banks that can be disabled would be more, but the number of misses and thus the performance penalty would also increase. So, to find the best trade-off between the power saving and the performance penalty, we estimate various thresholds and find that 0.5% is a suitable value for $C_{th}$. We also consider the maximum powered-off banks ($N_{off}$) for this policy to reduce the miss rate, and we set it 16 empirically.

### 4.4 Power-on Policy

Due to the changes in cache behavior in different execution time periods, we have to control cache banks all the time. Since LLC size decreases when some banks are power-off, increasing miss rates may reduce performance. Therefore, we always use entire counters of power-off banks to determine whether the program needs larger space.

At first, we write back the valid data of banks with the power-off opportunity before disabling it. In other words, if a dirty cache line is in compressed form, there is no need to write back because data is available in a compressed form during power-off. At the same time, the dirty cache line in the uncompressed style should be sent to the write-buffer. Therefore, there is no decompression delay for write-back processing. Furthermore, we reduce the write-back penalty as fewer writes would occupy the write buffer. Then, the power-on policy can be developed and discussed considering two cases: turning on each bank individually; turning on several banks simultaneously. First, when the sum of a bank's compressed lines and invalid lines is less than a predefined threshold ($Th_{ind}$) compared to its accesses, this bank is powered on. Second, since all accesses to the powered-off banks are not missed, we consider the number of misses. To detect the number of misses for powered-off banks, we don't need to add extra counters, and it can be calculated by the $C_A$ and $C_C$ of powered-off banks ($C_A$-$C_C$). Therefore, when the number of misses in powered-off cache banks is high (>$Th_t$), Nonbanks are powered on for performance improvement. The $N_{on}$ is dependent on the number of powered-off cache banks in an interval and should be set to half of this number. Actually, in each interval, the number of extra misses is checked; if it is more than That, half of the powered-off banks are powered on until the number of misses is set to less than That. After analyzing the different threshold impacts, we detect that a suitable value is 0.7% for $TH_{ind}$ and 1% for $Th_c$.

EECache presents an equation that determines how many powered-off banks should be active. In contrast, our power-on policy, along with turning on several banks, considers turning on each bank individually. Therefore, the proposed method determines which bank is the best choice for shutting on.

### 4.5 Migration Policy

We use the data migration policy to reduce the miss rate in powered–off banks and guarantee the data coherency. At the same time, we notice that migration may increase misses when the evicted victims in the active banks are reused later. Therefore, we limited migration based on a policy used to power off banks. Since the number of accesses to the uncompressed lines shut down by nonuniformity accesses in an interval is low, we don't use migration, and all valid uncompressed lines are discarded (M=1). While if nonuniformity accesses power off a bank in consecutive intervals, valid uncompressed cache lines of that bank are migrated to other active banks (M=0). The LRU lines in active banks are replaced during the migration.

EECache uses a hotness threshold to identify hot cache slices where data migration is done. At the same time, we notice that migration may increase misses when the evicted victims in the active banks are reused later. Since the number of accesses to a bank's uncompressed lines shut down by non-uniformity accesses in an interval is low, we don't use migration, and all valid uncompressed lines are discarded. If non-uniformity accesses power off a bank in consecutive intervals, valid uncompressed cache lines of that bank are migrated to other active banks. The LRU lines in active banks are replaced during the migration. Therefore, our work has a lower migration overhead than EECache because a big part of the cache is available in the power-off state.

## 5 EVALUATION METHODOLOGY

### 5.1 Experimental Platform

TABLE 2
SPECIFICATION OF THE EMBEDDED CMP CONFIGURATION

| Component | Description |
|---|---|
| Number of Cores | 16, 4×4 mesh |
| Core Configuration | Alpha21264, 3GHz, 45nm |
| Private Cache per each Core | SRAM, four-way, 64B line size, 32KB per core |
| On-chip Memory | Shared; 8MB total (64 SRAM banks, each of which has 128KB capacity) |
| Network Router | 2-stage wormhole-switched, virtual channel flow control, 2VCs per port, 5flits buffer depth, eight flits per a data packet, one flit per address packet, 16-byte in each flit |

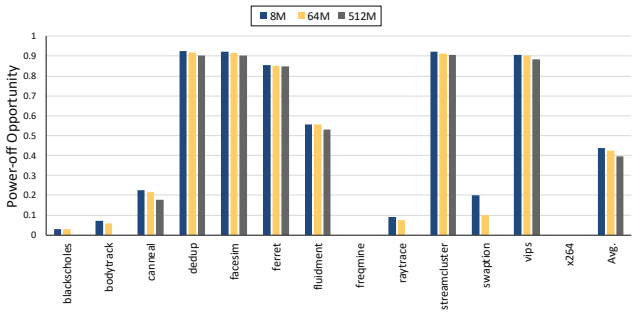

Fig. 9. Power-off opportunity. Of NFVCache for three different timing intervals.

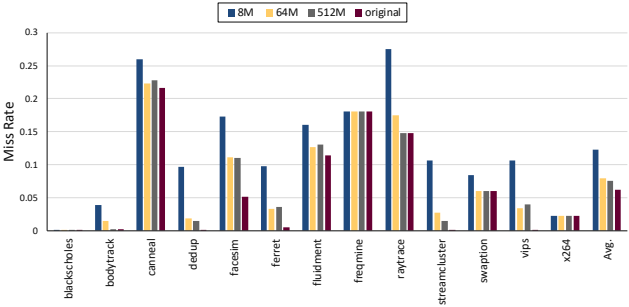

Fig. 10. Compare the miss rate of NFVCache for three timing intervals to the original cache.

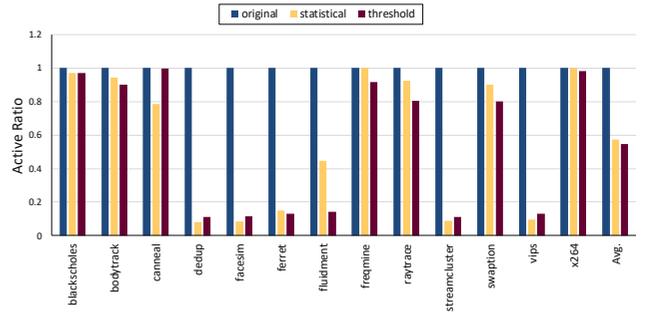

Fig. 11. Active ratio comparison for two power-off policies.

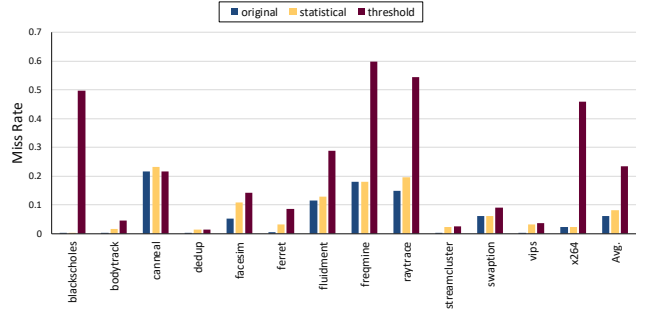

Fig. 12. Miss rate comparison for two power-off policies.

We use GEM5 [22] full system simulator in 32 nm technology to set up the primary system platform to evaluate the proposed method. Ruby is enabled to model a memory system and interconnect network. We model a multicore architecture with 16 cores that form a 2Dmesh topology, as shown in Fig. 8 (c). The main parameters of simulation configuration are listed in Table 2. Each core has a private L1 instruction and data cache. All cores share an L2 cache divided into 64 banks, and each bank has an area of 3.65 mm². The size of a core tile in the core layer is 3.5 mm² estimated by McPAT [23] and CACTI 6.0 [24]. The power of each bank is estimated by periodically interrupting the simulation timing of the gem5 and moving its output data into McPAT. Then to calculate the bank's temperature, the determined power of the McPAT feeds into Hotspot [38]. We employed HotSpot version 5.2 as a grid-based thermal modeling tool for temperature estimation. In our floorplan, we consider only banks as thermal blocks. We use the PARSEC and SPEC benchmarks. They are the most significant and standard candidates used in scientific studies to present CMP performance. We use simlarge as the input set in the PARSEC benchmark's initial analysis and final evaluation. Our observation was showed that our results are relatively independent of the size of input data sets. (simsmall, simmedium). For Parsec benchmarks, we executed 1 billion instructions for simlarge input set starting from the Region of Interest (ROI). We consider a bare-metal execution of applications in the done analysis.

### 5.2 Experimental Results

We first illustrate experimental results to find the best timing interval for checking binary counters defined in Section 4.3. Due to that FVCache is a general form than NIZCache, we chose an NFVCache architecture to determine the best timing interval. The power-off opportunity and the miss rate are shown in Fig. 9 and Fig. 10 as factors to select the best interval for Parsec benchmarks. We compare three timing intervals ranging from 8M cycles up to 512M cycles to a traditional cache for each application in each figure. Due to being powered-off all banks in a conventional cache, the power-off opportunity is 0% and we remove this case from Fig. 10. The power-off opportunity increases for shorter intervals and therefore significantly reduces leakage power, while the miss rate crucially increases. Generally, these figures suggest that intervals of 64M cycles have the best trend for energy saving and performance.

The circuits we add to the LLC cache to achieve our goal impose energy overhead in the LLC cache, in addition to delay overhead. Therefore, the minimal consists of two parts: first, the leakage energy due to the set of counters and the power-gating circuit, second dynamic energy due to extra cache misses and migration. However, we significantly have energy saving in the last level cache by minimizing the number of the power-on banks and reducing the number of writes and reads in LLC for zero-valued data in NIZCache and frequent values in NFVCache. We analyze energy consumption for NIZCache and NFVCache, considering these energies.

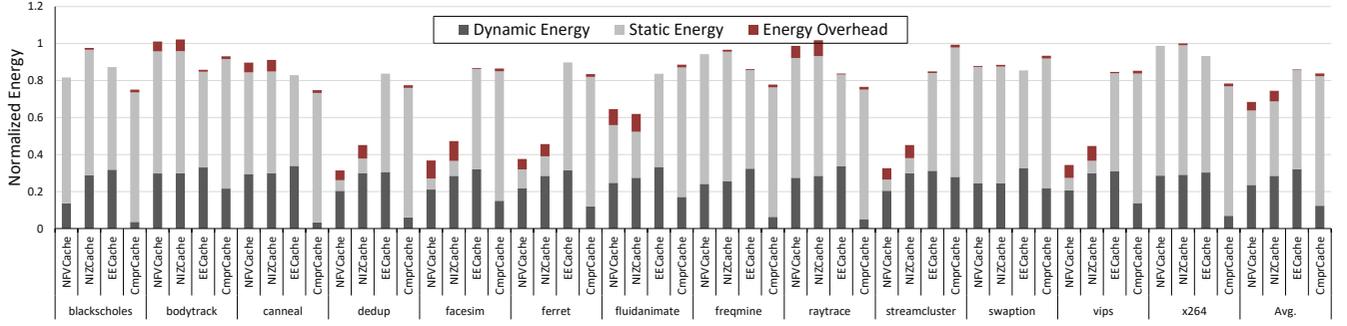

Fig. 13. Comparison of normalized energy consumption NFVCache, NIZCache, EECache, and CmprCache based on the baseline under execution of PARSEC benchmarks.

The active ratio and the miss rate are shown in Fig. 11 and Fig. 12 for NIZCache using two power-off policies alone: statistic-based (statistical) and threshold-based (threshold) policies. In the threshold approach, we find the cache banks with the non-uniform distribution by defining threshold instead of using standard deviation and mean; other policies, i.e., the migration policy and power-on policy, are the same with statistical approach. Due to being powered on all banks in a traditional cache (original), the active ratio is 100% in Fig. 11. As shown in Fig. 12, since the statistic-based power-off policy is decided individually based on program behavior, it increases the miss rate less than the threshold-based power-off policy. Therefore, NIZCache can benefit from the strength of this statistical information with less performance degradation. NIZCache, with a statistical approach on average, yields 43% energy reduction in LLC while increasing 22% miss rate.

We evaluate our cache effectiveness compared to a state-of-the-art slice-based exploiting design (EECache) [12]. We also compare our work with the state-of-the-art compression-based cache (CmprCache) design proposed in [27]. This coding method is similar to our work in encoding the frequent values using limited weight codes (1-LWC). As you can see in Fig. 13, this technique decreases dynamic energy more than our method because the length of FV is four bytes. However, this short length of frequent value increases the overhead of the compression/decompression circuit, leading to performance degradation. Fig. 13 shows the breakdown of the energy consumption into "static energy," "dynamic energy," and "energy overhead" (the leakage energy of the extra hardware and the dynamic energy of additional misses and migration) for CmprCache, EECache, NIZCache, and NFVCache. This figure shows that ache reduces energy consumption by about 12% and 15% on average compared with CmprCache and EECache, respectively. On the other hand, the NFVCache reduces energy consumption by about 22%, 25%, and 8% on average compared with CmprCache, EECache, and NIZCache, respectively.

As shown in Figure 13, in memory-intensive applications such as dedup, facesim, fluidanimate, streamcluster, and ferret, the amount of off-chip energy consumption due to increasing cache miss rate is more than in computation-intensive applications. Therefore, in computation-intensive applications such as blackscholes, bodytrack, swaption, and x264, the amount of static energy is more than dynamic energy because most of the instructions are done in the cores and do not need to access to the memory systems and most of the time, memory systems are not active and consume static energy. In memory-intensive applications such as dedup, facesim, fluidanimate, streamcluster, and ferret, static energy is less than dynamic energy because most instructions interact with memory systems. Figure 13 shows that the energy overhead in memory-intensive applications is more than in other applications.

On the other hand, we reduce energy consumption in interconnects of NFVCache by turning on only 1/16 wires in about 25% of communication between layers identified as frequent value for PARSEC benchmarks. Moreover, we evaluate the energy of each TSV in this section due to the presence of the capacitance crosstalk in interconnects. This energy consists of wire transition energy ($E^T$) and inter-wire transition energy ($E^C$). The total energy is the sum of these two energies for each TSV. In inter-wire transition energy caused by crosstalk, we consider only one neighbor from each category of TSVs, the north (i+1) and northwest (i+2) of TSV i, to avoid extra summation.

$$E = \sum_{i=1}^{N} E_i^T + \sum_{i=1}^{N} E_i^C = C_L V_{DD}^2 \sum_{i=1}^{N} P(trans_i) + \sum_{i=1}^{N} E_i^C \quad (4)$$

$$E_i^C = C_c V_{DD}^2 P(V_i(t) \neq V_{i+1}(t)) * E_t(i, i+1) + C_d V_{DD}^2 P(V_i(t) \neq V_{i+2}(t)) * E_t(i, i+2) \quad (5)$$

In the above equations, $P(trans_i)$ and $P(V_i(t) \neq V_{i+1}(t))$ denote the probability of transition in TSV $i$ and the probability of different voltage between TSVs $i$ and $i+1$, respectively. $E_t(i, i+2) \propto 0$ if $V_i(t) = V_{i+2}(t)$ otherwise $E_t(i, i+2) \propto$ number of transition in $(V_i(t^-), V_{i+2}(t^-)) \rightarrow (V_i(t), V_{i+2}(t))$. We reduce the probability of transition by limiting weight in codeword.

$$P(trans_i) = P(FV)P(trans_{FV}) + (1 - P(FV))P(trans_{uncoded}) \quad (6)$$

In above equation, $P(FV)$ and $P(trans_{FV})$ denote the probability of the FV presence and the probability of tran-sition in TSVs consist of 1-LWC coding, respectively. For uncoded case, the value of $P(trans_i)$ and $P(V_i(t) \neq V_{i+1}(t))$ are 0.5 because there is no compression for the input data. While in this method, we reduce the probability of transition by restricting the weight (the number of 1s) in a codeword. Therefore, due to the probability of transition, we can obtain $P(V_i(t) \neq V_{i+1}(t))$ and $E_t(i, i + 1)$. We suppose $P(V_i(t) \neq V_{i+1}(t)) = 1/2$ in the first state. And then there are two cases to evaluate the probability of inequality after the transition. In the first case, a transition passes only in one of two wires that have same initial voltage. And in the second case with different initial voltages, a transition passes in both wires or there is no transition. Similarly, we can calculate the value of $E_t(i, i + 1)$ by the following equation:

$$E_t(i, i + 1) = \sum_{T=0}^{2} T * P(T) = 2\left(1 - P(trans_i)\right)P(trans_i) + 2(P(trans_i))^2 \quad (7)$$

TABLE 3
MULTI PROGRAMMED WORKLOADS

| Test program suite | Benchmarks |
|---|---|
| Memory Bounded set1 (MB1) | zeusmp(2), libquantum(2), lbm(2), GemsFDTD(2), art(4), swim(4) |
| Memory Bounded set2 (MB2) | zeusmp(3), libquantum(3), lbm(3), GemsFDTD(3), art(2), swim(2) |
| Memory Bounded set3 (MB3) | zeusmp(4), libquantum(4), lbm(4), GemsFDTD(4) |
| Medium set1 (MD1) | mcf(2), sphinx3(2), bzip2(2), calculix(2), leslie3d(2), gcc(2), cactusADM, milc, omnetpp, wupwise |
| Medium set2 (MD2) | mcf(3), sphinx3(3), leslie3d(2), gcc(2), cactusADM(2), milc(2), omnetpp(2) |
| Medium set3 (MD3) | mcf(2), sphinx3(2), bzip2, calculix, leslie3d(2), gcc(2), cactusADM(2), milc(2), omnetpp(2) |
| Computation Bounded set1 (CB1) | parser(2), applu(2), face_rec(2), equake(2), astar(2), hmmer(2), bzip2(2), calculix(2) |
| Computation Bounded set2 (CB2) | parser(2), applu(2), face_rec(2), equake(2), astar(2), hmmer(2), bzip2, calculix, mpeg_dec(2) |
| Computation Bounded set3 (CB3) | parser(2), applu(2), face_rec(2), equake(2), astar(2), hmmer(2), mpeg_dec(4) |
| Mixed set1 (Mix1) | sphinx3, mcf, astar(2), hmmer(2), gamess(2), perlbench(2),gromacs(2), tonto(2), gcc, leslie3d |
| Mixed set2 (Mix2) | sphinx3(2), mcf, astar(2), hmmer, gamess(2), perlbench(2),soplex, gromacs, gcc(2), leslie3d(2) |
| Mixed set3 (Mix3) | sphinx3(2), mcf(2), astar, hmmer, gamess, perlbench, soplex(2), gcc(3), leslie3d(3) |

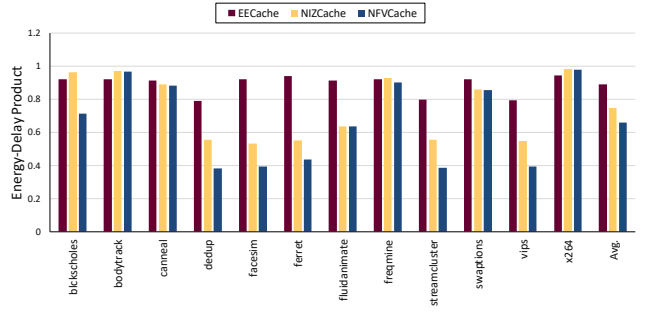

Fig. 14. Comparison of energy-delay product results in compared to EECache under multithreaded workloads.

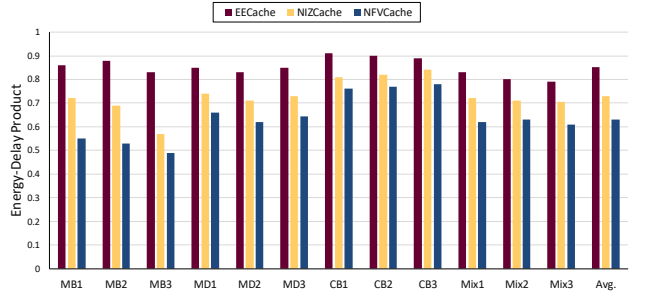

Fig. 15. Comparison of energy-delay product results in compared to EECache under multiprogrammed workloads.

Based on two above equations, the parameter values of $P(trans_i)$ and $E_t(i, i + 1)$ for NFVCache are 0.378 and 0.756, respectively. Now if we suppose that $C_c = 5.54C_L$ [20] then we figure out that $C_d = 1.385C_L$. Consequently, due to Equation 4 and 5, the energy consumption of uncoded cache and NFVCache for a TSV equal to $14.1C_L V_{DD}^2$, $8.1657C_L V_{DD}^2$, respectively. We can see NFVCache is 72% less than uncoded cases. Therefore, the interconnect energy in capacitive crosstalk improves in NFVCache.

**Overhead.** Since we place all counters related to banks in the cache controller, we do not need to gather all the counter information from distributed cache. So, it does not require data transfer through the on-chip network, resulting in time and energy overhead. Therefore, we need three significant parts to evaluate the total overhead of our proposed design:

• In NIZCache design, a bit for each 64-byte line of the last level cache is used to determine zero values. Since we use a 32MB LLC, the storage overhead for this bit is 64KB that is 0.19% of LLC. While in the NFVCache design, we use sixteen 2KB tables in each tile to determine frequent values in addition to 64KB for an extra bit (32KB+64KB=96KB) that is 0.29% of LLC.

• A set of counters to find the power-off opportunities of cache banks. So, since we use three 12-bit counters for each bank, the storage overhead for this part is (3*12bit*64(bank) = 288B) 0.00086% of LLC, in addition to using a 12-bit counter for each bank to count misses in each bank used in the power-on policy of NFVCache (12*64+288B=384B) that is 0.00114% of LLC.

• An array of Gated-VDD circuitry to turn on/off the cache banks. Since we use bank-based shutdowns, the total area overhead is 2.3% of LLC for the two designs.

Thus, the total overhead is estimated to be 2.5% and 2.6% of the last level cache for NIZCache and NFVCache, respectively, if we consider that the area overhead is proportional to the storage.

Calculating the mean and standard deviation for only 64 dataem's delay 12 bits also does not create the system since they perform in the background. Moreover, the hardware implementation of the calculation of standard deviation and mean, as shown in Fig. 16, consumes 152mW of power, and it does not provide a significant overhead in our proposed method (area=0.330mm²). First, most previous works like EECache used adder/subtractor, multiplier, and divider modules to determine their decision. Second, we have a lot of free space in the first layer to implement this architecture in a tile

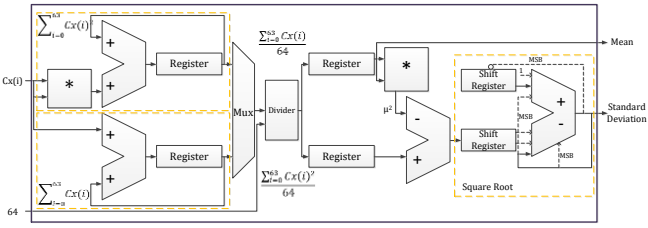

Fig. 16. Basic hardware architecture for computing the mean and the standard deviation.

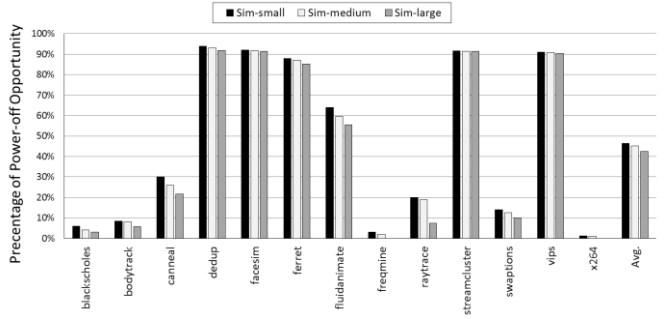

Fig. 18. Sensitivity to total LLC cache capacity and turn-off opportunity.

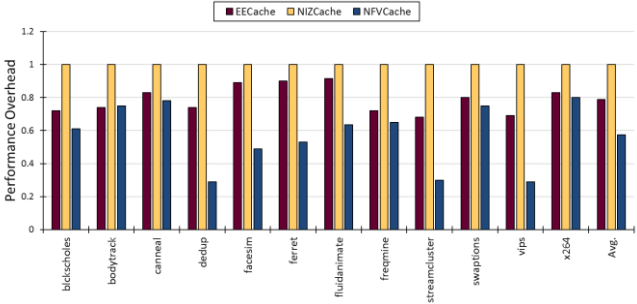

Fig. 17. Performance overhead in three different methods normalized to NIZCache

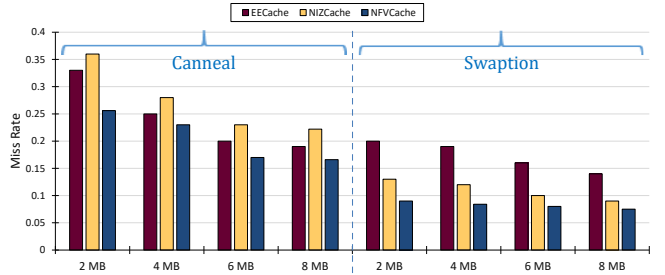

Fig. 19. Comparison of miss rate with increasing the on-chip memory from 2MB to 8MB for swaption and canneal.

due to the core size and the cache bank size.

The coding circuit adds one cycle to write access of our LLC cache to evaluate delay overhead. Also, it consumes less than 40μW of power per column even when actively used on every cycle. Synopsys Design Compiler acquired our measurements of the delay and the power of coding and decoding circuits. However, since we used a bit for each cache line instead of the sampling array, the hardware overhead for two architectures is slightly more significant than the prior work [12]. Furthermore, NFVCache increases the delay overhead due to searching in the FV table for each access.

**Sensitivity Analysis**

**Sensitivity to multi-programmed and multithreaded workloads:** In this section, we analyze the sensitivity of our proposed method to multithreaded and multiprogrammed workloads. We choose the energy-delay product (EDP) parameter for this sensitivity analysis to represent energy consumption and performance best. We consider multiprogrammed workloads consisting of 16 applications for performing our experiments in this part. The applications are selected from the SPEC2000/2006 benchmark suites [37]. Based on the memory demand intensity of benchmark applications, we classified them into three groups: memory bounded, medium, and computation bounded benchmarks. We generated a range of workloads (combinations of 16 benchmarks), as summarized in Table 3. Note that, the number in parentheses is the number of instances. Moreover, we use multithreaded workloads selected from PARSEC for performing the multithreaded experiments the same as previous experiments. Under a random mapping approach, applications/threads are randomly mapped to the cores. Figs. 14 and 15 show the results of EDP for multithreaded and multi-programmed workloads, respectively.

We compare our proposed method to EECache in this subsection. As shown in Fig. 14, NFVCache improves EDP by about 26% on average compared to EECache. In addition, NFVCache improves EDP by about 11% on average compared to NIZCache. NVFCache focuses on reducing interconnect energy consumption as one contributor to uncore energy consumption in parallel with cache banks more than NIZCache. EECache considers cache banks to minimize energy consumption while NFVCache and NIZCache reduce the energy consumption of cache banks and interconnects with the lowest performance degradation. As shown in Fig. 15, NIZCache improves EDP by about 14% on average compared to EECache. Furthermore, NFVCache improves EDP by about 13% compared to NIZCache. CB test program suites are computation-intensive suites. Therefore, cores in the core layer work at maximum frequency and do not work with uncore, and the effect of proposed methods cannot be seen on them compared to memory-intensive workloads.

Regarding this point, that EECache method relies on the static energy and the active footprints of the LLC and does not consider the information of data type and distribution them in banks; therefore, losing performance can be seen in workloads with high non-uniformity zero lines over cache banks such as *freqmine*, *swaption*, and *Mix2* more than other applications in Figs. 14 and 15. These results represent that our design can improve power saving with a slight increase in overhead. In PARSEC applications, interthread communication is deficient. Therefore, the EDP improvement of NFVCache for these workloads compared to NIZCache is less than multi-programmed SPEC2006.

As shown in Fig. 17, the performance overhead has been calculated for EECache, and NFVCache normalized to NIZCache. This figure shows that NFVCache shows 21% and performance improvement compared to NIZ-

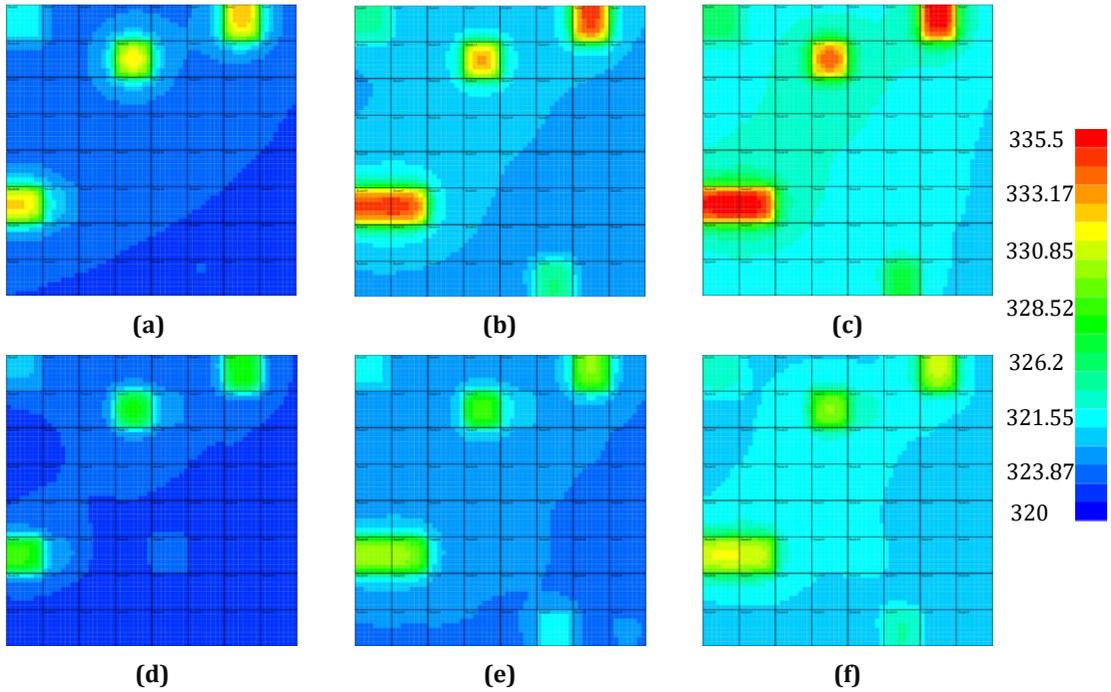

Fig. 20. Heat maps of NFVCache (a, d), NIZCache (b, e), and Baseline (c, f) for streamcluster and facesim benchmarks, respectively.

Cache. Moreover, EECache shows 10.3% and performance improvement compared to NIZCache.

**Sensitivity to the inputs size of mapped applications:** As mentioned before, all of the PARSEC benchmarks were executed for simlarge input sets in prior evaluations. It is noticeable that PARSEC benchmarks have multiple input sizes, simsmall, simmedium, and simlarge. Active last level cache capacities change execution time according to the working sets of PARSEC benchmarks defined by their input sizes. Therefore, it would be helpful to see a sensitivity study concerning the total previous level cache capacity in the proposed architecture to understand what fraction of turn-off opportunities were due to asymmetric utilization vs. low working set to cache size ratio. We show this sensitivity study in Fig. 18.

We considered two representatives for the performance reduction of the cache for different applications, swaption from computation-intensive applications and canneal from memory-intensive applications. As shown in this Fig. 19, the performance improves by increasing the on-chip memory size in canneal. Increasing the on-chip memory size for the swaption application has performance improvement but not as much as canneal.

Fig. 20 shows the temperature distribution in the cache layer of the system for two applications in Baseline (conventional cache) and proposed architectures. Fig. 20(a)-(c) show thermal maps of NFVCache (a), NIZCache (b), and Baseline (c) for *streamcluster* benchmark, respectively. The results show that the maximum temperature in NFVCache is reduced by 3.39 K compared to Baseline. The temperature of some banks is reduced significantly; for instance, bank 41 is reduced by 12.09 K concerning this bank in Baseline. Fig. 20(d)-(f) show similarly thermal maps of NFVCache (d), NIZCache (e), and Baseline (f) for *facesim* benchmark, respectively. Baseline architecture under execution of all the PARSEC applications operates around and upper the critical temperature. The proposed architecture underperformance of blackscholes, bodytrack, and freqmine as computation-intensive applications work near but under the required temperature. Compared with the Baseline architecture does not violate the maximum temperature. Under execution of facesim, fluidanimate, ferret, dedup, and streamcluster as memory-intensive workloads, the proposed architectures work because the proposed techniques have been proposed for the situation that uncore components have an impressive contribution.

## 6 RELATED WORK

For energy-efficient CMP architecting, Esmailzadeh et al. [1] showed that fundamental performance limitations stem from the processor cores. They ignored the power impact of "uncore" components such as the cache hierarchy, memory subsystem, and on-chip interconnection. In this paper, we focus on LLCs as one of the most critical components in uncore architectures, as with technology scaling and increasing number of cores on a chip in CMPs, the number of "uncore" components will increase, and hence they will further eat into the power budget, reducing speedups. This section reviews some prior techniques related to LLC's dynamic and static power reduction.

To reduce the dynamic power, some techniques use compression. The dynamic zero compression [25] is an old technique that uses zero values to save cache energy. This technique monitors the written value and compresses the zero value into a flag. Dusser proposed Zero-Content Augmented (ZCA) cache at al. [14] that uses an extra specialized cache (ZC) to store null cache lines. Due to the high spatial locality of null cache lines, a single tag can use for several null lines. Zero-value Canceling Cache [26] also uses zero loads that consist of about 18% of the total dynamic loads to respond quickly and improve performance. Moreover, several works that exploited the zero bits in the written values have been introduced to improve power [39][40] and reliability [41]. Jung et al. [39] use zero-valued data of cache lines to reduce the write and the read energy of the STT-RAM cache by putting an extra bit for every word. However, although their proposed cache has compression and decompression circuits with low latencies, it is limited for some applications with many null cache lines. But our work is more general than this limitation, and we use zero values as part of our proposal to reduce power in LLC and interconnection networks. On the other hand, Yazdanshenas et al. [27] proposed a coding scheme for Non-Volatile Memory (NVM) LLC based on value locality to make writes less, but actually, the authors reduced dynamic power while we concentrate not only dynamic energy but also static energy.

Some prior studies proposed several cache power-off strategies that reduce the leakage power and turn off unused cache parts. Cache decay [28] is a famous cache leakage-saving technique that employs power-gating at granular cache lines. This work proposed a decay-based approach to powered off cache lines that its data cannot be reused after a per-set number of cycles. While this fine-grained power-gating strategy may cause significant hardware overhead compared to bank-level power-gating. Recently, Cheng et al. [12] represented a slice-based cache organization to shut down with minimal overhead. They proposed a run-time design called EE-Cache that uses three parameters, including utilization, hotness, and writeback of dirty data to shut down slices of the LLC for energy saving. Arima et al. [29] presented a power management technique called Immediate Sleep (IS) to architect an energy-aware LLC. IS uses a next-access predictor algorithm to shut down subarrays of a STT_MRAM cache by power-gating. In their algorithm, if the subsequent access is not critical in performance, the subarray related to this access is immediately shut down. In this work, we use power-gating at the cache bank granularity. Power-gating or gate-Vdd [42] and drowsy cache [43] are two fundamental circuit-level methods to minimize the leakage power in traditional SRAM caches. On the other hand, emerging memory technologies with low leakage power are proposed to replace conventional SRAM cache. Jadidi et al. [30] used the non-uniformity of write accesses to hybridize SRAM and NVM technologies. This approach either uses a lot of specific thresholds to distinguish parts of the cache or requires high hardware overhead to predict cache behavior. This paper uses statistical information instead of specific thresholds to determine banks in the non-uniform cache architecture.

To address leakage power that is a primary source of total NoC power, darkNoC [31] was recently proposed, which uses multi-Vt based on multiple routers. This NoC architecture consists of multiple network layers with architecturally homogenous routers, and each layer is optimized to operate in an especial voltage-frequency (VF) range at design time. At each time, only one network layer is active, and the other network layers are deactivated. Therefore, all routers have the same VF simultaneously. In the power-gated NoC designs, ShuttleNoC [32] has proposed a new NoC architecture to enforce optimal power efficiency. We see that almost all these previous studies are concerned with designing router microarchitectures for achieving power efficiency in NoC. At the same time, in this paper, we investigate the role of data encoding to reduce power in NoC instead of developing sophisticated router architectures. The data encoding scheme is another method that was employed to reduce the link power dissipation [33] [34] [45-48]. NoΔ compression [35] was proposed to reduce traffic in Network on Chip and save energy by lowering the network load. Delta Compression requires additional logic in the NI. This other logic consists of vector addition, subtraction, and comparison operations. Therefore, this method has modest constant hardware overhead and implementation complexity, while our coding has negligible overhead.

## 7 CONCLUSION

This paper proposes two power management mechanisms for last-level cache, exploiting the power-off opportunities provided by the non-uniform distribution of cache accesses, compressed cache lines, and invalid lines. At every power management time interval, we use the average and standard deviation of valid uncompressed accesses across banks in that interval and previous interval to turn off cold banks. These architectures solve the static power problem in SRAM caches. Results show that the first architecture, NIZ-Cache based on zero values, improves the energy-delay product by about 17% and 14% on average compared to the EECache approach under multithreaded and multi-programmed workloads execution, respectively. And second design, NFVCache based on frequent valthe use, improves the energy-delay product by about 26% and 25% on average compared to the EECache approach under multmulti programmedulti-programmed workloads execution. Our proposed ideas are limited to applying the SRAM cache, but also our metrics can be added to most methods such as EECache. We can use the mentioned metrics in a hybrid SRAM/STT-RAM cache to enable large 3D stacked shared LLCs that the two used compression codings also can decrease write energy and improve the life of NVMs integrated into the LLC.